\documentstyle[prb,floats,eqsecnum,aps,epsf]{revtex}
\begin{document}
\draft

\twocolumn[\hsize\textwidth\columnwidth\hsize\csname@twocolumnfalse%
\endcsname

\title{ 
Quantized Thermal Transport in the Fractional Quantum Hall Effect}
 
\author{C.L. Kane}
 
\address{Department of Physics, University of Pennsylvania\\
Philadelphia, Pennsylvania 19104
}
\author{Matthew P.A. Fisher}
\address{Institute for Theoretical Physics, University of California\\
Santa Barbara, CA 93106-4030
}

\date{\today}

\maketitle

\begin{abstract}
We analyze thermal transport in the fractional quantum Hall effect (FQHE),
employing a Luttinger liquid model of edge states.
Impurity mediated inter-channel scattering events are incorporated
in a hydrodynamic description of heat and charge transport.
The thermal Hall conductance, $K_H$, is shown to provide a new and universal
characterization of the FQHE state, and reveals non-trivial information
about the edge structure.
The Lorenz ratio between thermal and electrical
Hall conductances {\it violates} the free-electron Wiedemann-Franz law,
and for some fractional states is predicted to be 
{\it negative}.
We argue that thermal transport may provide a unique way to detect
the presence of the elusive upstream propagating modes,
predicted for fractions such as $\nu=2/3$ and $\nu=3/5$.

\end{abstract}
]

\section{Introduction}
The connection between quantized electrical transport and
the microscopic structure of edge states is of fundamental importance to
the quantum Hall effect\cite{Halperin}.
The Hall conductance is related to the additional edge
current $J$ which flows when the chemical potential $\mu$
of the edge is raised,
\begin{equation}
G_H = e{\partial J\over{\partial\mu}}.  
\end{equation}
In the integer quantum Hall effect, the edge states
consist of a single non-interacting electron mode
for each full Landau level.
Due to the cancellation between velocity and 1d density of states,
the contribution of each mode to the Hall conductance has
the quantized value, $e^2/h$.

In addition to charge, energy is also transported by quantum
Hall edge states.  At temperatures well below the quantum Hall gap,
the energy moving along the edge cannot easily
escape, since there are no bulk current
carrying electronic excitations.
When the top and bottom edges of a Hall bar are
at different temperatures
a thermal transport current will then flow.  
This gives rise to the thermal analogue of the Hall effect,
known as the Leduc-Righi effect\cite{Landau}.
One can define
a thermal Hall conductance, analogous to (1.1),
\begin{equation}
K_H = {\partial J_Q \over{ \partial T}},
\end{equation}
where $J_Q$ is the thermal current carried by the edge modes.
For the free electron edge modes in the integer quantum
Hall effect, the edge velocity also cancels for 
the thermal current \cite{Imry}  and 
each mode contributes a quantized amount to $K_H$, of magnitude
\begin{equation}
K_H = {\pi^2\over 3} {k_B^2\over h} T.
\end{equation}
The thermal $K_H$ and electrical $G_H$
Hall conductances are thus simply related by the Wiedemann-Franz law for
free electrons.  A similar quantization of the thermal conductance
occurs for a quantum point contact.

In the fractional quantum Hall effect (FQHE) the edge modes are
no longer free electron-like, but rather are chiral
Luttinger liquids\cite{Wen}.  The charge carried by
these modes contributes
to the electrical Hall conductance, giving an appropriately
quantized fractional value.
But as in the IQHE, one anticipates
that the edge modes will also dominate the
transport of heat at low temperatures.
In this paper we study thermal transport in the
fractional quantum Hall effect.  
By employing a chiral
Luttinger liquid model of the edge states, we find that
the FQHE edge modes also contribute a quantized thermal Hall conductance.
But in contrast to the IQHE, $K_H$ is no longer related
to the Hall conductance $G_H$, via the Wiedemann-Franz law.
Rather, $K_H$ provides an independent quantized characterization
of the FQHE state.  
In fact, the quantized thermal
Hall conductance is a new and universal
property of the quantum Hall state, in some ways as fundamental
as the electrical Hall conductance, although of course much more
difficult to measure.
But if measured,  $K_H$ would provide a nontrivial test
of microscopic edge state theories, as we elucidate below.
A similar violation of the Wiedemann-Franz law occurs in a 
non-chiral Luttinger liquid\cite{KFthermal}

For hierarchical FQHE states, multiple
propagating modes on a given edge are predicted\cite{Wen}.
But even more intriguing is the prediction that
for certain filling fractions, 
such as $\nu=2/3$ and $\nu=3/5$, some of the chiral edge channels
propagate in the ``wrong" direction -- opposite to that
of the classical skipping orbits specified by
the sign of the magnetic field\cite{Wen,MacDonald}.
Such ``upstream" modes have not yet been detected
experimentally\cite{Ashoori}, presumably due to 
to effects of edge state equilibration. 
In this paper we show that these upstream modes can have a profound effect
on the thermal transport.  Specifically, for the fraction $\nu=3/5$
the thermal Hall conductance $K_H$ is predicted to
be {\it negative} - of opposite sign to the
electrical Hall conductance, $G_H$.  The upstream modes actually
dominate in the thermal transport, and carry heat in the
direction opposite to the charge transport.
For $\nu =2/3$, on the other hand, we predict that the thermal Hall
conductance vanishes,
due to a cancellation between up and down stream modes.  
Rather than being carried ballistically,
the heat transport along the edge is predicted to be diffusive,
leading to a non-vanishing thermal Hall {\it conductivity}.
These predictions are robust, being valid in the presence
of equilibration processes, due say to edge impurity scattering.
Thus, thermal 
transport measurements may provide a unique way to 
to establish the existence of the elusive upstream moving channels.

The outline of our paper is as follows.
In Section II we consider thermal transport for an ideal, impurity free,
edge. In Section III we generalize the 
theory to include disorder mediated inter-channel scattering
events.  Specifically, we formulate a hydrodynamic
theory which is valid on time scales long compared
to the equilibration times.  
In this regime, there remains only a single
hydrodynamic charge mode, and a single heat mode,
even for a multi-channel edge.
Section IV is devoted to a brief discussion of experimental
implications.  Specific geometries are considered, which
should allow for measurement of edge heat transport.

\section{Clean Edge}

Consider the edge of a clean fractional quantum Hall 
effect state at filling $\nu$,
in equilibrium at chemical potential $\mu$ and temperature $T$.
In this section we compute both the electrical and thermal
Hall conductances, using the Luttinger liquid model.
At the $n th$ level of the Haldane-Halperin hierarchy\cite{HaldaneHalperin},
$n$ chiral Luttinger liquid edge modes are expected\cite{Wen}, 
described by the Hamiltonian,
\begin{equation}
{\cal H}_0 = \pi \sum_{ij} N_i V_{ij} N_j.
\end{equation}
Here $N_i$ are one-dimensional densities in the $n$ channels,
and the
$V_{ij}$ are non-universal interactions,
determined by the edge confining potential and 
inter-channel Coulomb forces.
The densities $N_i$ obey the Kac Moody commutation algebra
\begin{equation}
[ N_i(x), N_j(x') ] = 
{i\over {2\pi}} {\sf K}^{-1}_{ij} \partial_x \delta(x-x') ,
\end{equation}
where the $n$ by $n$ universal matrix ${\sf K}$ characterizes 
the topological order in the bulk quantum Hall fluid\cite{WenZee,Read}.
The electrical charge carried by each mode is specified
by a vector $T_i$, so 
that the total edge charge density is
$n_\rho = \sum_{i} T_i N_i$.
The filling factor is related to ${\sf K}_{ij}$ and $T_i$ via
$\nu = \sum_{ij} T_i  {\sf K}_{ij}^{-1} T_j$.
Specific representations of the matrix ${\sf K}$ can be
found in Ref. \onlinecite{WenZee}.
   
The electrical Hall conductance which follows
from this model is appropriately quantized.
To see this, first note that the
operator equations of motion imply
a continuity equation, $\partial_t n_\rho + \partial_x
J_\rho$, with a conserved charge current given by
$J_\rho = \sum_i T_i ({\sf K}^{-1} V)_{ij} N_j$.
Now, add a constant potential $\Phi$ which couples
to the total  
charge in the Hamiltonian:  ${\cal H}_\Phi = -n_\rho \Phi$.
Upon minimization of the Hamiltonian, this results
in a transport current $J_\rho = G_H \Phi$ with,
\begin{equation}
G_H = \nu {e^2\over h}.
\end{equation}

Before analyzing the transport of heat, it is convenient
to first re-cast (2.1) and (2.2) into a diagonal form,
with a transformation  $N_i = \sum_j \Lambda_{ij} n_j$.
As shown in Ref. \onlinecite{KFsun},
$\Lambda$ can be chosen so that 
both matrices $K$ and $V$ are brought into diagonal form:
$(\Lambda^T {\sf K} \Lambda)_{ij} =
\eta_i \delta_{ij}$ and $(\Lambda^T V \Lambda)_{ij} = 
v_i \delta_{ij}$.  
In this new basis, the transformed Hamiltonian becomes
\begin{equation}
{\cal H}_0 = \pi \sum_i  v_i n_i^2,
\end{equation}
with the new densities satisfying
\begin{equation}
[ n_i(x),  n_j(x')] = {i\over {2\pi}} \eta_i \delta_{ij}
\partial_x \delta(x-x').
\end{equation}
In this basis, the model describes
$n$ independent modes which propagate with a speed
$v_i$ in a direction specified by $\eta_i= \pm 1$.
The number of up and down stream modes is a universal property of 
the matrix ${\sf K}$.
The charge associated with each mode $n_i$ is 
$t_i = \sum_j \Lambda_{ij}^T T_j$.

Thermal currents can now be readily extracted.
Each channel describes independent chiral density modes 
which propagate at speed $v_i$ in the $\eta_i$ direction,
and have energy-momentum dispersion
$E_i(q) = \hbar  v_i q$.
At temperature $T$, each mode is 
thermally populated with a Bose distribution function,
$b(E/k_BT)$.
The resulting thermal current is simply,
\begin{equation}
J_Q = \sum_i \eta_i  v_i n_{Qi},
\end{equation}
where $n_{Qi}$ denotes the energy density in channel $i$.
For an edge in local equilibrium at temperature $T$, this can be
expressed as a sum over all the modes:
\begin{equation}
n_{Qi} =  \int {dq\over 2\pi} E_i(q)  b({E_i(q)\over{ k_B T}}) = 
{1\over  v_i} {\pi^2\over 6} {k_B^2\over h} T^2  .
\end{equation} 
Upon insertion of (2.7) into (2.6),
the non-universal velocities $ v_i$ are seen to cancel,
giving a thermal Hall conductance,
\begin{equation}
K_H = {\partial J_Q\over{\partial T}} =  \nu_Q {\pi^2\over 3}{k_B^2\over h}T .
\end{equation}
The coefficient 
$\nu_Q = \sum_i \eta_i$ is the difference between the
number of upstream and downstream channels.

Upon combining with (2.3), the Lorenz ratio can be expressed as:
\begin{equation}
L = {K_H\over {T G_H}} = \left({\nu_Q \over \nu}\right) L_0 ,
\end{equation}
where $L_0 = (\pi^2/3)(k_B/e)^2$ is the free-electron value.
Notice that each mode contributes the same amount to the
thermal conductance, whereas the ``charges" $t_i$ implicitly
enter the electrical conductance, since $\nu = \sum_i \eta_i t_i^2$.
This leads to a Lorenz ratio for the FQHE which violates 
the Wiedemann-Franz law.
Like the Hall conductance $\nu$,
the thermal coefficient $\nu_Q$ is a robust and universal
topological quantity, characterizing the Hall state.

When all of the edge channels move in the same direction, 
as shown for $\nu=2$ in Fig. 1a, $K_H$ is simply a 
measure of the total number of channels.   
But when channels move in both directions,
there is an exact cancellation between the
contribution of the up and downstream modes to the thermal
Hall conductance.
This leads to some striking predictions.
For $\nu=3/5$ the form of the $K$ matrix indicates
that there are three modes, two of which move upstream,
as sketched in Fig. 1b.
This implies a thermal Hall conductance which is {\it negative} - 
opposite in sign to the electrical Hall effect.  
For $\nu=2/3$ (Fig. 1c)
there is one upstream channel and one downstream channel.  Thus,
$K_H = 0$: in equilibrium there is no net heat flow along the edge.
In the next section we argue that these results are robust, and will
survive the presence of edge impurity scattering which serve to
equilibrate the various edge modes. 

\begin{figure}
\epsfxsize=2.8in
\centerline{ \epsffile{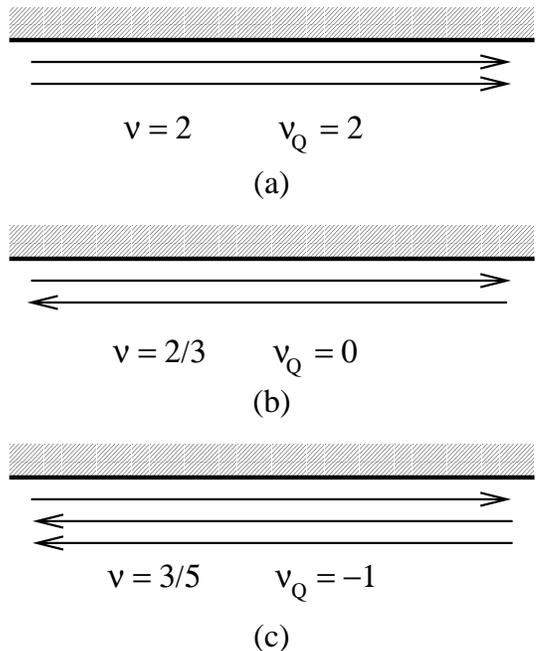} }
\caption{
Schematic representation of edge states for several different
quantum Hall fluids at filling factor $\nu$.  
The integer $\nu_Q$, which specifies
the difference between the number of downstream and upstream
propagating modes, determines the sign and magnitude of the thermal
Hall conductance.
}
\end{figure}

\section{Hydrodynamic Theory}

A central feature of the Luttinger liquid theory
of FQHE edges, is the existence of upstream charge carrying modes.
Unfortunately, these upstream modes have {\it not} been
detected experimentally, despite experiments by Ashoori et. al.
\cite{Ashoori} designed explicitly
to observe them.  
We believe the resolution of this problem lies
in the assumption of a clean edge\cite{KFnu23,KFcontacts}.
The clean edge model analyzed in the previous section,
contains an artificially
high symmetry.  In particular, 
the charge in each of the $n$ different channels is independently 
conserved.  The existence of $n$ conserved charges implies
$n$ hydrodynamic modes - those described by (2.4) - some of which
propagate upstream.  

Impurities which are inevitably
present near the edges of any real sample,
will destroy this symmetry.
One expects that the only remaining conserved charge is the total
electrical charge density on the edge.
This would imply the presence of only a single long-lived mode,
analogous
to zero sound in a Fermi liquid.
Actually, the situation is somewhat more complicated.
As shown in detail in Ref. \onlinecite{KFsun,KFnu23}, 
even with impurity scattering present there are
$n-1$ {\it other} conserved ``charges" at zero temperature,
arising from symmetries associated with channel inter-change.
In most cases, these other ``charges" are neutral with
respect to electric charge, and
so do not contribute
directly to electrical transport.  These $n-1$ neutral modes
are analogous to quasiparticle excitations in a Fermi liquid.
At non-zero temperatures, the ``charge" associated with
these other $n-1$ modes is no longer conserved, and they decay away by
scattering.  Thus, at finite temperature, the only remaining
conserved charge is indeed total electric charge.  

In this paper we focus exclusively on the hydrodynamic regime - at
frequency scales below these relevant decay rates.
In this regime, we expect only a single 
propagating hydrodynamic mode, associated with total charge conservation,
which propagates downstream.
All upstream charge motion is associated with non-conserved charges,
and decays away.  Observation of upstream charge transport
requires a ``mesoscopic" experiment in which the sample is
smaller than the edge equilibration length.

In the absence of electron-phonon coupling, the electronic energy is
also conserved at the edge.  Thus, there should be an additional
hydrodynamic mode associated with thermal transport.  
Provided the coupling to the phonons is sufficiently weak, this can
be a long lived mode.  In this section we develop 
a simple hydrodynamic theory of charge and heat transport,
based on a Boltzmann transport equation.
In addition to the hydrodynamic charge mode (zero sound), we identify the 
single hydrodynamic mode describing the flow of (conserved) 
energy.  This hydrodynamic heat mode leads to a
quantized thermal conductance, as in the previous section.
Moreover, when $\nu_Q <0$, this hydrodynamic mode will
be shown to propagate upstream -
in the opposite direction
to the hydrodynamic charge mode.

Consider first a hydrodynamic description of charge
propagation.
To formulate a Boltzmann transport theory, we first
consider the ``collisionless" regime for a clean edge, 
described by the de-coupled Hamiltonian (2.4).
The response of the system to a spatially and temporally
varying potential $\Phi$ which couples to the total
charge density, $n_\rho = \sum_i t_i n_i$,
may be obtained from the operator equations of motion as\cite{KFcontacts}:
\begin{equation}
(\partial_t + \eta_i  v_i \partial_x)  n_i
= \eta_i  t_i \partial_x \Phi.
\end{equation}

The transport equation (3.1) conserves the charge
in each mode.  It is the analog of the
collisionless Boltzmann equation for a Fermi liquid.
Impurity scattering 
will destroy this conservation and lead to a flow of 
charge and energy between the channels.   This may
be incorporated into the transport equation 
by including a collision term\cite{KFsun,KFcontacts}.  
This Boltzmann description of transport is only valid provided the
collisions - or tunneling events - occur incoherently.
As discussed above, this implies that the system is in the
hydrodynamic
regime, at scales longer than equilibration
lengths and times.
The appropriate linearized transport equation takes the form,
\begin{equation}
(\partial_t + \eta_i  v_i \partial_x)  n_i
+ \sum_j M_{ij}  n_j = \eta_i t_i \partial_x \delta\Phi.
\end{equation}
The matrix $M_{ij}$ is in general complicated and
depends on the detailed nature of the inter-channel scattering.
However, $M$ must obey two constraints
imposed by conservation laws.
(i) In equilibrium there must no net steady state flow
of charge between
the channels.  The equilibrium densities can be extracted by
taking a constant potential, $\Phi = \mu$, the chemical potential.
Then
$ n_i =  v_i^{-1}  t_i \delta \mu$.
This implies the condition,
\begin{equation}
\sum_j M_{ij}  v_j^{-1}  t_j = 0.
\end{equation}
(ii)  The total charge on the edge must be conserved, even
out of equilibrium.
Since the charge flowing out of channel $i$ is 
$\sum_j  t_i M_{ij}  n_j$,
this implies
\begin{equation}
\sum_i t_i  M_{ij}  = 0.
\end{equation}

Since the transport equation (3.2) is linear,
it may be solved by Fourier transforming, to obtain
an eigenvalue problem, with eigenvalues $\omega(q)$. 
Most of the eigenvalues will correspond to solutions
which decay exponentially in time.
However, the two constraints, (i) and (ii) 
above, guarantee that there is one low
frequency mode, with $\omega \propto q$ as $q \rightarrow 0$.
Specifically, consider a solution to (3.2) of the form
\begin{equation}
 n_i(x,t) =  
n_\rho { t_i  v_i^{-1}\over \sum_i t_i^2  v_i^{-1}} 
e^{i(qx-\omega t) }.
\end{equation}
For $q,\omega\rightarrow 0$ this corresponds to ``local
equilibrium" with a slowly varying charge
density $n_\rho e^{i(qx-\omega t) }$. 
Since there is only a single conservation law - total electric charge - 
we expect this to be the only low frequency solution.  
Inserting (3.5) into (3.2), multiplying by $t_i$ and summing on i, 
allows the total electric charge $n_\rho$ to
be related to $\Phi$.   
The current-charge response function, defined by
$J_\rho = \Pi_\rho(q,\omega) \Phi$, readily follows,
\begin{equation}
\Pi_\rho(q,\omega) = 
G_H {v_\rho q\over{v_\rho q - \omega - i\eta}},
\end{equation}
with,
\begin{equation}
v_\rho = \nu \left(\sum_{i}  t_i^2   v_i^{-1}\right)^{-1},
\end{equation}
and $G_H$ given in (2.3).
The response function has a single pole, describing a
single hydrodynamic charge mode, which propagates downstream
at velocity
$v_\rho$.  This is in contrast to the clean edge,
which has $n$ propagating modes, which may also move
upstream.
As expected,   
the quantized Hall conductance is given by the $q\rightarrow 0,\omega=0$ limit 
of the response function.

Consider now a similar analysis for the thermal transport.
The transport equation for the
energy density in each channel, given a spatially and
temporally varying temperature, takes the form,
\begin{equation}
(\partial_t + \eta_i  v_i \partial_x)  n_{iQ}
+ \sum_j M^Q_{ij}  n_{jQ}
 = \eta_i {\pi^2\over 3}T\partial_x \delta T.
\end{equation}
Here we have included a collision term which
describes the flow of heat between
the channels.
Conservation of energy demands that the matrix
$M^Q$ must satisfy
\begin{equation}
\sum_i  M^Q_{ij} = \sum_j M^Q_{ij}  v_j^{-1} = 0,
\end{equation}
the thermal analogs of conditions (i) and (ii) above.
The appropriate generalization of (3.5),
which describes a low frequency thermal mode corresponding
to local thermal equilibrium, is,
\begin{equation}
 n_{Qi}(x,t) = n_Q {v_i^{-1} \over { \sum_i v_i^{-1}}}
e^{i(qx - \omega t)}  .
\end{equation}
The thermal response function can then be obtained, as before,
employing the thermal Boltzmann equation (3.8).  This gives, 
\begin{equation}
\Pi_Q(q,\omega) = K_H {v_Q q\over{v_Q q - \omega - i\eta}},
\end{equation}
where $K_H$ is the quantized thermal Hall conductance given in (2.8),
and the velocity of the heat mode is
\begin{equation}
v_Q = \nu_Q \left(\sum_{i} v_i^{-1}\right)^{-1}.
\end{equation}
Notice that
this velocity can be of either sign,
depending on $\nu_Q$. 
Thus for $\nu=3/5$, the hydrodynamic heat mode flows upstream
in the opposite direction of the charge.   

For $\nu=2/3$ the thermal Hall conductance vanishes,
and a more detailed analysis of (3.8) is required.
This reveals that
the thermal transport is diffusive, rather than ballistic,
with a response function given by,
\begin{equation}
\Pi_Q(q,\omega) = C { \omega q D_Q  \over{D_Q q^2 - i\omega }},
\end{equation}
where $C= (\pi^2/3)T \sum_i  v_i^{-1}$ is the edge heat
capacity.  Although the thermal conductance vanishes,
there is a finite thermal {\it conductivity} $\kappa = C D_Q$.
The value of the thermal diffusion constant $D_Q$ depends on the details of the 
scattering matrix $M^Q_{ij}$.

\section{Experimental Implications}

As shown above, the
thermal Hall conductance, $K_H$, 
contains important new information about the structure of the
edge excitations in the FQHE.  In particular, the sign of $K_H$ is sensitive
to the presence of edge modes which propagate ``upstream".
In this section we briefly discuss the feasibility
of measuring the Hall thermal conductance.
We suggest a particular geometry which
should at least enable a measurement of the
sign of $K_H$.

To extract $K_H$ requires measuring the thermal current carried by the edge
excitations.  
Although this is clearly much more challenging than measuring
charge transport, a recent
experiment by Molenkamp et. al.\cite{Molenkamp} has demonstrated
the feasibility of measuring thermal transport in mesoscopic structures.
Specifically, in this experiment the
thermal conductance of a quantum point contact (in zero magnetic field)
was extracted.  The trick was using {\it additional}
point contacts
as ``thermometers", to measure the local temperature
of the electron gas on either side of the point contact.
The additional point contacts were biased on the edge
of a step between two plateaus, so that they would have a large, temperature
independent thermopower of order $40 \mu V/K$\cite{Streda}. 
Then, by measuring the voltage across these additional
point contact thermometers the
local temperature change was extracted.
In Molenkamp's experiment, the thermal current was estimated
from the change in temperature by estimating
the heat capacity of the electron gas.
This allowed for a determination
of the thermal conductance and Peltier coefficient
of the point contact which agreed favorably with
theoretical expectations.

\begin{figure}
\epsfxsize=2.8in
\centerline{ 
\epsffile{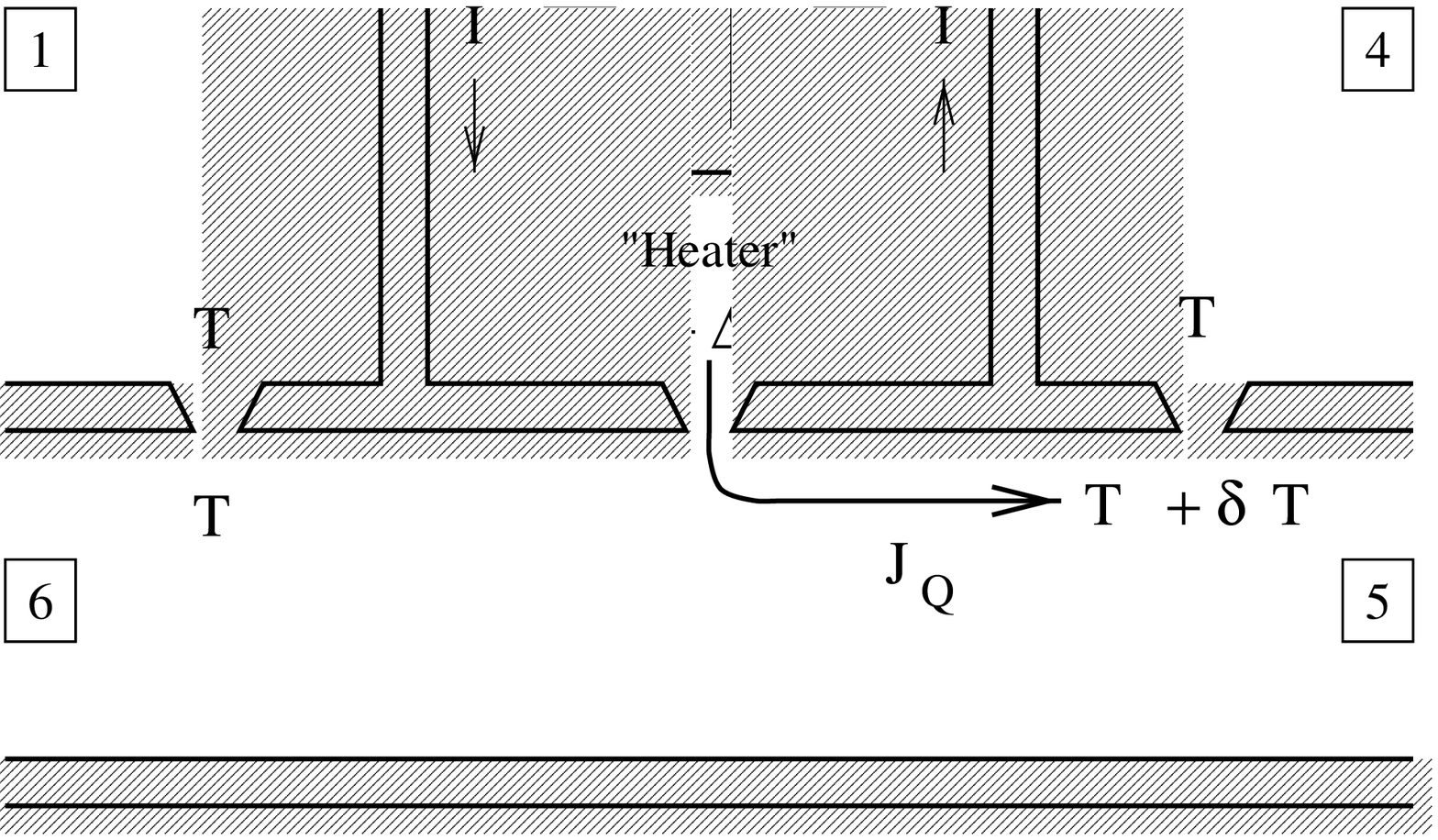} 
}
\caption{Proposed geometry for measurement of the thermal Hall
conductance.  The heat generated by passing
a current between contacts 2 and 3,
is detected by two additional point contacts, upstream (1-6)
and downstream (4-5) along the edge.
}
\end{figure}

It should be possible to adapt this technique
to measure the thermal transport of quantum Hall 
edge states.  
As a concrete example, consider the geometry
sketched in Fig. 2.  
As in the experiment by Molenkamp et. al.,
the sample can be heated locally by driving a small  
electric current through the electron gas.
Specifically, a current between contacts 2 and 3 (see Figure)
would locally heat the edge of the quantum Hall fluid.
Alternatively, it might be possible to
heat the edge directly by coupling in a local
RF probe\cite{Leo}.
This local heating will 
be carried away by the edge states, either raising the
temperature of the downstream or upstream edge, depending
on the sign of $K_H$.
This temperature change can
then be detected by measuring the voltages
across the point contact thermometers.

While this measurement might not be
suitable to extract the
magnitude of $K_H$, it
should be adequate to determine
the direction of heat propagation - and hence
the {\it sign} of $K_H$. 
For this, one need only detect an {\it asymmetry}
in the temperature change 
of the upstream and downstream edges.  
For integer quantum Hall states, and fractional states
at filling
factors such as $\nu =1/3, 2/5$,  
the heat should flow downstream, resulting
in a temperature increase at the downstream thermometer only.
On the other hand, for filling $\nu = 3/5$, since $K_H$
is negative, heat flows ``upstream", so the temperature
increase should be detectable at the upstream thermometer.
For $\nu = 2/3$, the heat is predicted to diffuse along the edge.  The
temperature increases should then be the same at the up and downstream
thermometers, but smaller in magnitude. 
  
On sufficiently long length scales, the edge states will
thermally 
equilibrate with the phonons in the substrate.
It is thus crucial that the heater and thermometer are closer
together than the electron-phonon thermal equilibration length.
While this equilibration length has not been measured, we
expect it to be quite long.  
Energy relaxation rates have been measured for 2 dimensional
electron gases in GaAs\cite{Dzurak,Ma} in zero magnetic field. 
At $1^\circ K$ relaxation times of order $20 ns$
are found, which suggests an equilibration length
of upwards of $100 \mu m$.
At low temperatures, the interaction between electrons
and the lattice in GaAs is dominated by the piezoelectric
coupling to acoustic phonons and the relaxation rate
decreases as the temperature is lowered\cite{Ma,Ridley}.
Thus, we suspect that lattice thermalization will not be a 
problem in the proposed edge state experiment.

\section {Conclusion}

We have shown that the thermal conductance 
of a quantum Hall edge state is universal and quantized: 
$K_H = \nu_Q K_0$ with $K_0 = (\pi^2/3)(k_B^2/h) T$.
The integer $\nu_Q$ specifies
the difference between the number of downstream and upstream
edge modes.
Moreover, the quantization of $K_H$ was shown to be robust, 
valid in the presence of interactions and impurity scattering
at the edge.   We also expect the quantization to hold
for an edge with a slowly varying confinement potential, 
which may have channels in addition
to those required by the Luttinger liquid theory\cite{Chklovskii}.
Since the additional edge channels come in pairs - one upstream and one
downstream - they do not change $\nu_Q$.
On sufficiently long length scales, so that all edge
modes are equilibrated,
we predict only a single
hydrodynamic charge mode and a single heat mode.
These two hydrodynamic modes carry the charge and thermal currents, leading to the quantized conductances.

In the FQHE, the thermal Hall conductance $K_H$ contains additional information about the microscopic edge
structure, not present in the electrical
conductance.  For $\nu=3/5$, we predict that
$K_H$ is {\it negative},
due to the presence of ``upstream" propagating edge modes.
A measurement of the sign of $K_H$ would thus provide a critical
and nontrivial test of current edge state 
theories.

\acknowledgements

It is a pleasure to thank T. Heinzel, A.T. Johnson and
L. Kouwenhoven for helpful discussions.  We are grateful to the National
Science Foundation for support.  M.P.A.F has been supported by
grants PHY94--07194, DMR--9400142
and DMR-9528578.  C.L.K. has been supported by grant DMR 95-05425.

\end{document}